\documentclass{aastex}
\usepackage{spr-astr-addons}
\usepackage{epsfig}
\usepackage{graphicx}
\usepackage{color}

\begin{document}
%
\title{Observation of kink waves and their reconnection-like origin in solar spicules}

\shorttitle{kink waves in solar spicules}
\shortauthors{Ebadi et al.}

\author{H.~Ebadi, and M.~Ghiassi}
\affil{Astrophysics Department, Physics Faculty,
University of Tabriz, Tabriz, Iran\\
e-mail: \textcolor{blue}{hosseinebadi@tabrizu.ac.ir}}

\begin{abstract}
We analyze the time series of \mbox{Ca\,\textsc{ii}} H-line obtained from
\emph{Hinode}/SOT on the solar limb. We follow three cases of upwardly propagating kink waves along a spicule and
inverted Y-shaped structures at the cusp of it. The time-distance analysis shows
that the axis of spicule undergos quasi-periodic transverse displacement
at different heights from the photosphere. The mean period of transverse
displacement is $\sim\!\!175$ s and the mean amplitude is $1$ arc\,sec.
The oscillation periods are increasing linearly with height which may be counted as the
signature that the spicule is working as a low pass filter and allows only the low frequencies
to propagate towards higher heights. The oscillations amplitude is increasing with height
due to decrease in density. The phase speeds are increasing until some heights and then decreasing which may be related
to the small scale reconnection at the spicule basis.
We conclude that transversal displacement of spicules axis can be related to the propagation of kink waves along them.
Moreover, we observe signatures of small-scale magnetic reconnection at the cusp of spicules which may excite kink waves.

\end{abstract}


\section{Introduction}
\label{sec:intro}
Spicules are observable in H$_{\alpha}$,
D$_{3}$ and \mbox{Ca\,\textsc{ii}} H chromospheric lines. The general properties of them can be found
in some reviews \citep{bek68,ster00,zaq09}.
Observation of oscillations in solar spicules may be used as an indirect
evidence of energy transport from the photosphere towards the corona.
Transverse motion of spicule axis can be observed by both, spectroscopic and imaging observations.
The periodic Doppler shift of spectral lines have been observed from ground based coronagraphs \citep{nik67,Kukh2006,Tem2007}.
But Doppler shift oscillations with period of $\sim\!\!5$ min also have been observed on the Solar and Heliospheric
Observatory (\emph{SOHO}) by \citet{xia2005}.  Direct periodic displacement of spicule axes have been found by
imaging observations on Optical Solar Telescope (SOT) on \emph{Hinode\/} \citep{De2007,Kim2008,he2009,Ebadi2012}.

There are two different types of drivers in the highly dynamic photosphere:
oscillatory (e.g. p-modes) and impulsive (e.g. granulation and/or explosive events due to magnetic reconnection).
Both types of drivers may be responsible for the observed dynamical phenomena in upper atmosphere regions \citep{Tem2011}.
\citet{De2004} reported a synthesis of modeling and high-spatial-resolution observations of individual spicules.
In the upper chromosphere, p-modes are evanescent and cannot propagate upwards through the temperature minimum.
However, evanescence does not preclude significant tunneling of non-propagating wave energy into the hot chromosphere.
More importantly, they showed that the non-verticality of flux tubes dramatically increases tunneling, and can even lead to propagation of
p-modes. The leaked photosphere velocity signals steepen into shocks and the oscillatory wake behind them can led to spicule formation.
It should be noted that because of the presence of cut-off periods, only frequencies above the cut-off are propagating:
the frequencies below cut-off are evanescent. However, radiative relaxation of temperature perturbations is known to be
important and leads to the conclusion that all frequencies propagate \citep{Roberts1983}.

\citet{Tem2007} analyzed the consecutive height series of H$\alpha$ spectra in solar limb spicules and traced wave propagations
through the chromosphere. They used discrete Fourier transform analysis and detected Doppler-shifted oscillations.
They concluded that the oscillations can be caused by wave propagation in thin magnetic flux tubes. They suggested
the granulation as a possible source of the wave excitation. \citet{Ebadi2013} analyzed the time series of
H$\alpha$ line obtained from \emph{Hinode}/SOT on the solar limb. They used Wavelet analysis to show the transversal
oscillations of spicules axis. They found that the strong pulses may lead to
the quasi periodic rising of chromospheric plasma into the lower corona in the form of spicules.
The periodicity may result from the nonlinear wake behind the pulse in the
stratified atmosphere. \citet{Morton2012} based on SDO/AIA observations
detected a variety of periodic phenomena in conjunction with large solar jets.
Among other mechanisms that can be deduced from observations, they concluded that
only one reconnection event has occurred and we are seeing the response of the transition region
to a single velocity pulse. The rebound-shock model suggests
that a velocity pulse can cause the transition region to
generate a damped, (quasi-)periodic response. It has been demonstrated that the
waves in the lower solar atmosphere can drive reconnection to
produce small scale jet events with the same periodicity as the
exciting wave. On the other hand, waves generated by reconnection could also display these timescales.
\citet{Murawski2010,Murawski2011} performed numerical simulations of solar spicules and macro-spicules.
They studied the upward propagation of a localized velocity pulse that is initially launched below the transition region.
They showed that the strong initial pulse may lead to the quasi periodic rising of chromospheric material
into the lower corona in the form of spicules. The periodicity results from the nonlinear wake that is formed behind the pulse in the
stratified atmosphere. The superposition of rising and falling off plasma portions resembles the time sequence of single and double
(sometimes even triple) spicules, which is consistent with observational findings. They concluded that
the two-dimensional rebound shock model may explain the observed speed, width, and heights of type I spicules, as
well as observed multi-structural and bi-directional flows. The model also predicts the appearance of spicules with 3–5 min period
due to the consecutive shocks.

\citet{Kudoh1999} used the random nonlinear Alfv\'{e}nic pulses and
concluded that the transition region lifted up to more than $\sim5000$ km (i.e. the spicule produced).
They studied the case of random perturbation in the photosphere as a source
of Alfv\'{e}n waves in the context of spicule formation and
associated coronal heating.

\citet{he2009} based on \emph{Hinode}/SOT observations reported for the first time
the excitation of kink waves in spicules due to magnetic reconnection. They observed transversal displacement of
spicule axis which originated from the cusp of an inverted Y-shaped structure. They interpreted such structures as
evidences to the magnetic reconnection at the basis of spicules. \\
In the present work, we study the observed oscillations in the solar spicules through the data obtained from \emph{Hinode}.
We trace the spicule axis oscillations via time slice diagrams and deduce proportional periods at different heights.
The generation of kink waves in solar spicules due to small-scale magnetic reconnection will be discussed.

\section{Observations and image processing}
\label{sec:observations}
We used a time series of \mbox{Ca\,\textsc{ii}} H-line ($396.86$ nm) obtained on 18 October
2008 during 21:59 to 22:04 UT by the Solar Optical Telescope onboard \emph{Hinode\/} \citep{Tsu2008}. The spatial resolution reaches $0.2$ arc\,sec ($150$ km) and the pixel size is $0.109$ arc\,sec ($\sim\!\!80$ km) in the \mbox{Ca\,\textsc{ii}} H-line.  The time series has a cadence of $10$ seconds with an exposure time of $0.5$ seconds.  The position of $X$-center and $Y$-center of slot are, respectively, $0$ arc\,sec and $960$ arc\,sec, while, $X$-FOV and $Y$-FOV are $112$ arc\,sec and $56$ arc\,sec respectively.

We used the ``fgprep,'' ``fgrigidalign'' and ``madmax'' algorithms \citep{Shimizu2008,Koutchmy89}
to reduce the images spikes and jitter effect and to align time series and to enhance the finest structures, respectively.

\section{Results and discussions}
\begin{figure}[!h]
\epsscale{1.00}
\plotone{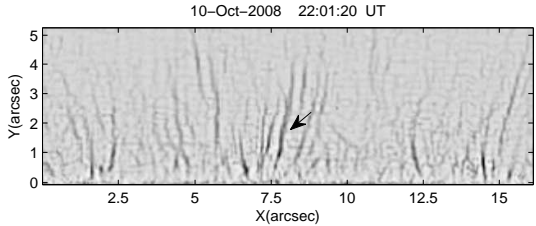}
\caption{The full view \mbox{Ca\,\textsc{ii}} H-line image of the solar north pole which contains the processed spicule.
The black arrow shows the studied spicule. \label{fig1}}
\end{figure}

\begin{figure*}
\centering
\includegraphics[width=20cm]{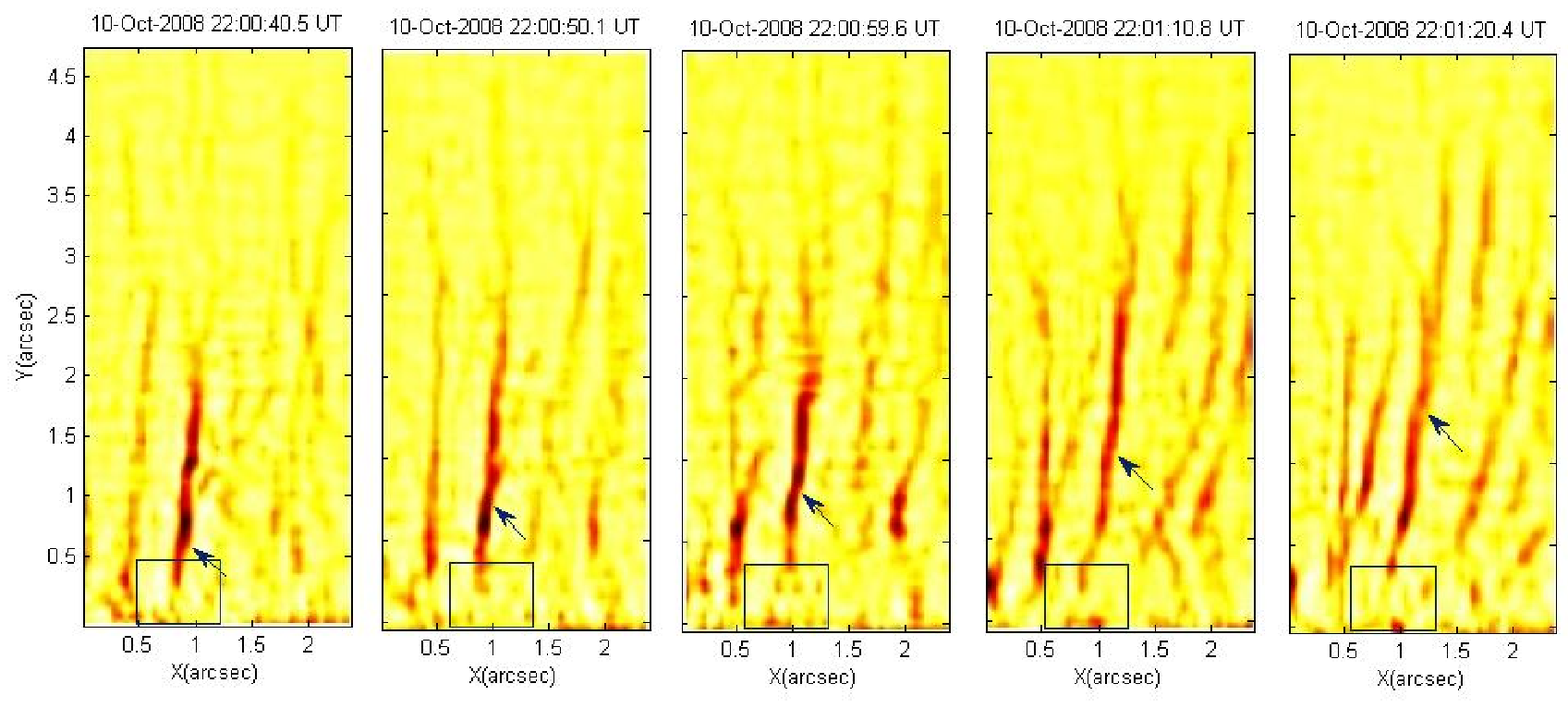}
\includegraphics[width=20cm]{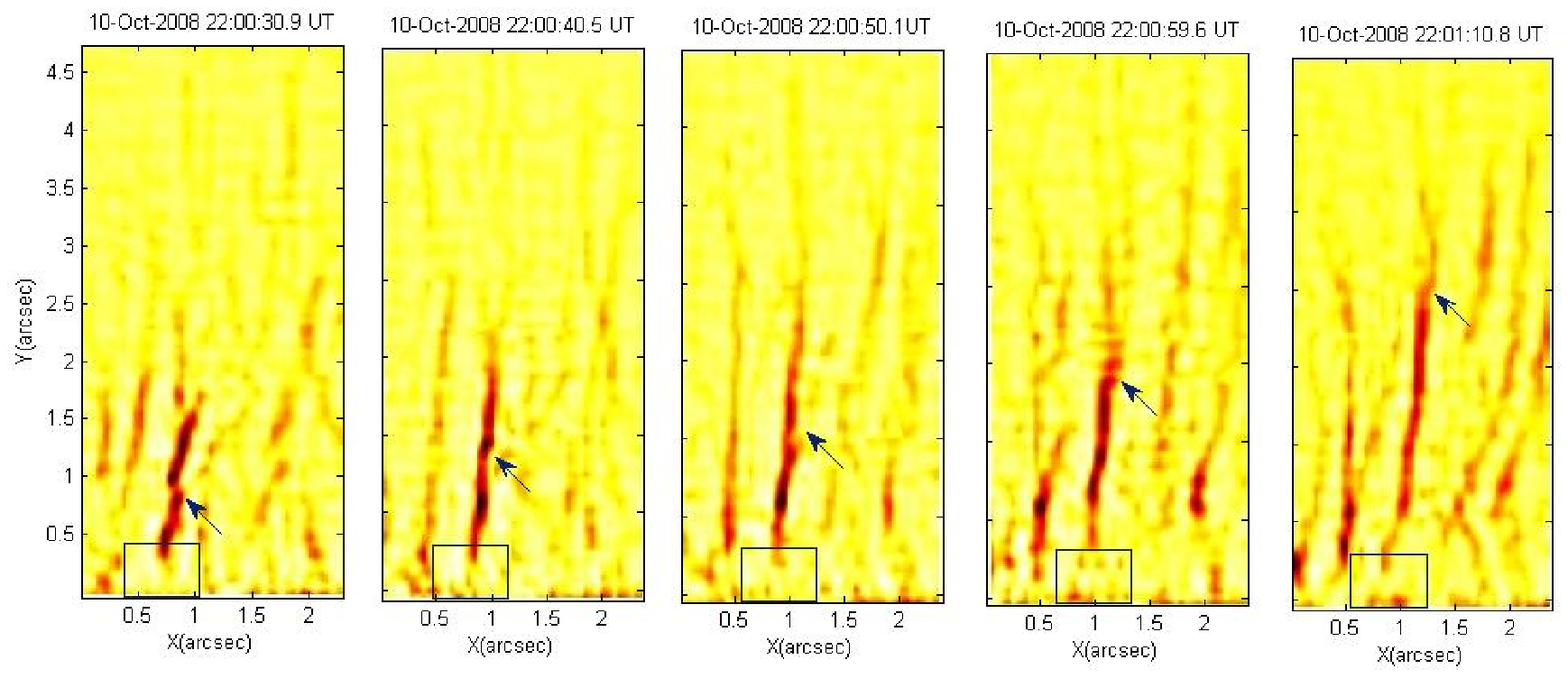}
\includegraphics[width=20cm]{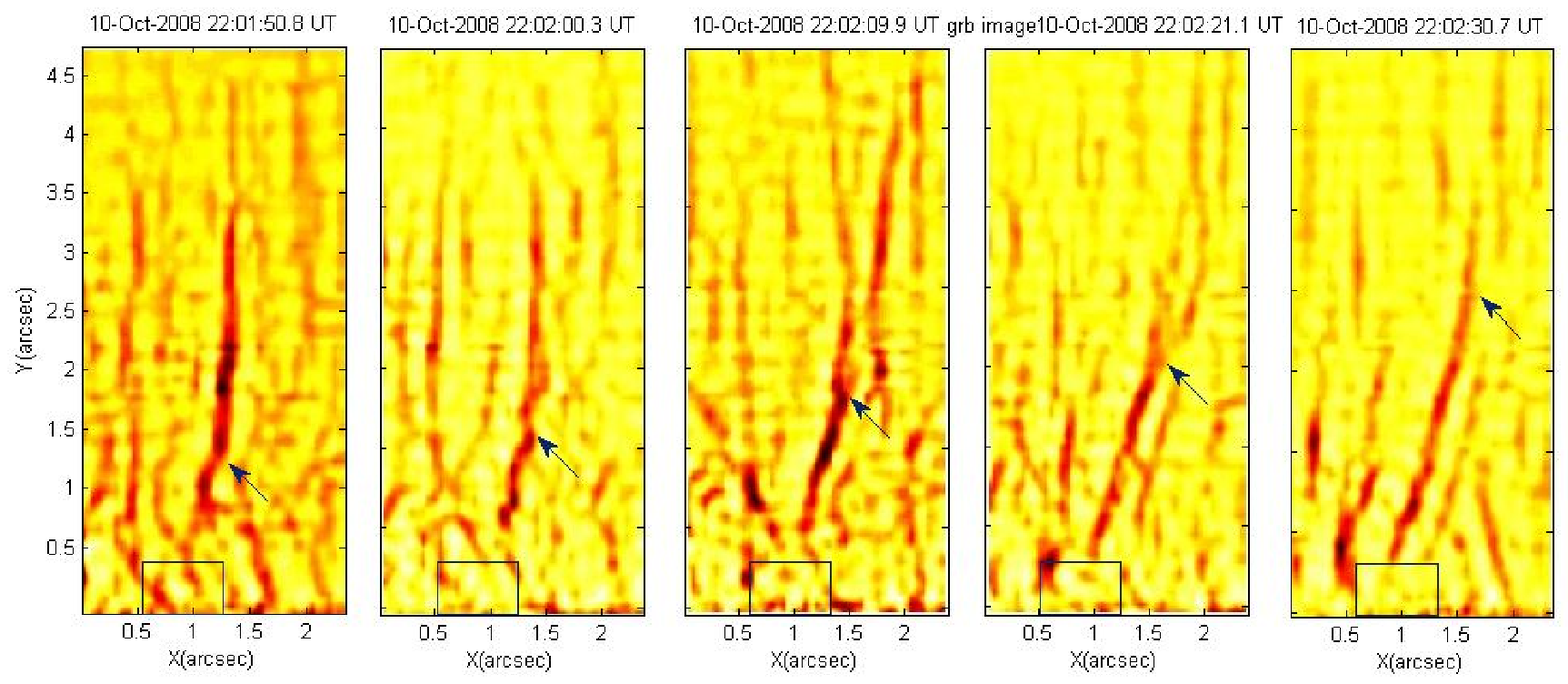}
\caption{Upwardly propagation of kink waves along a spicule in different times during its lifetime are presented.
The cusp of the inverted Y-shaped structures is showed by rectangular. \label{fig2}}
\end{figure*}

In Figure~\ref{fig1} we presented the full view \mbox{Ca\,\textsc{ii}} H-line image of the solar north pole which contains the studied spicule.
The processed spicule is indicated by a black arrow on this image. This image was taken by \emph{Hinode}/SOT telescope on $10$ Oct $2008$, $22:01:20$ UT.
 We used the time series images of this region and determined the average length, width, and lifetime
of the studied spicule as $\sim3000$ km, $\sim200$ km, and $\sim4$ min, respectively.
Figure~\ref{fig2} shows $5$ selected images of the time series which trace spicule dynamics.
In this figure we follow three different signatures of transversal oscillations of a unique spicule during its lifetime.
As is clear from this figure, the followed spicule is the same, but the sequence of time is different
from top to bottom. In other words, we found three cases of upwardly propagating kink wave along a spicule
in different times during its lifetime. These oscillations are showed by black arrows on the figure.
The inverted Y-shaped structures are seen at the spicule base which may correspond
to the magnetic reconnection origin of it \citep{he2009}.


\begin{figure*}[!h]
\epsscale{2.00} \plotone{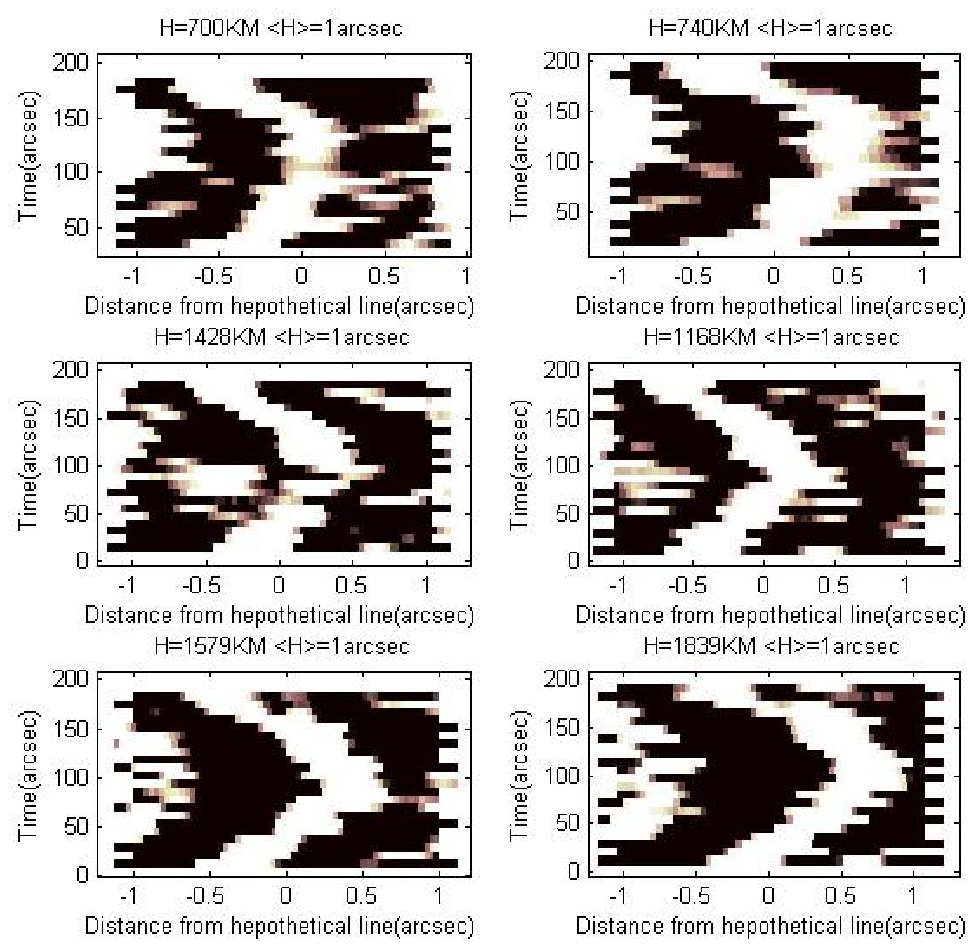} \caption{Time slice images of the time series
in different heights are shown. \label{fig3}}
\end{figure*}

Figure~\ref{fig3} shows the time slice diagrams performed at $6$ different
heights from the limb. Each cut is obtained by averaging over $9$ pixels along the spicule axis,
which corresponds to $1$ arc\,sec around each height. The clear quasi-periodic transverse
motion of the spicule axis is seen on the figure. The white regions represent the spicule at
particular height.  We clearly see that the spicule axis undergoes
the transverse oscillation at each height with nearly same periodicity.

\begin{figure}
\centering
\includegraphics[width=8cm]{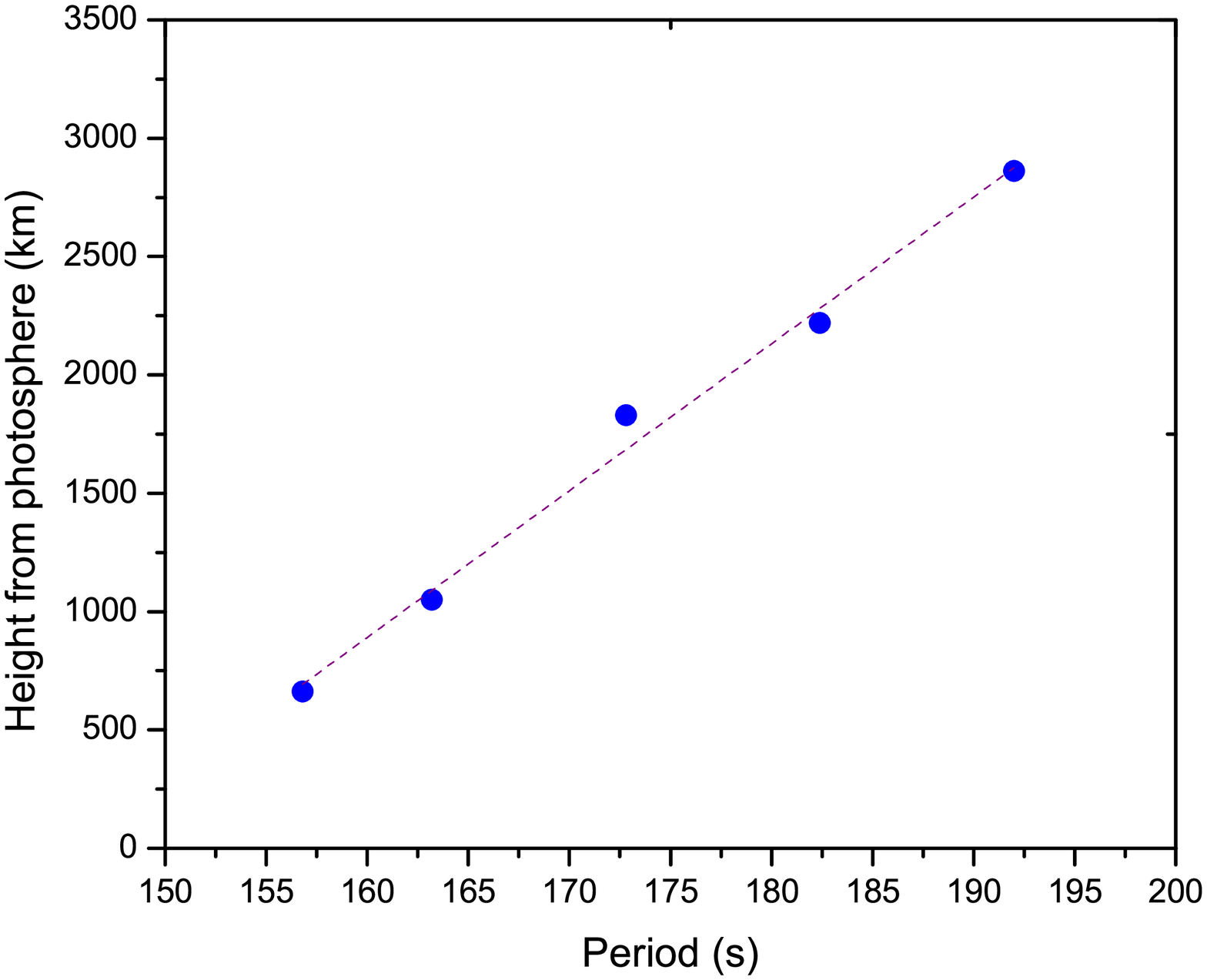}
\includegraphics[width=8cm]{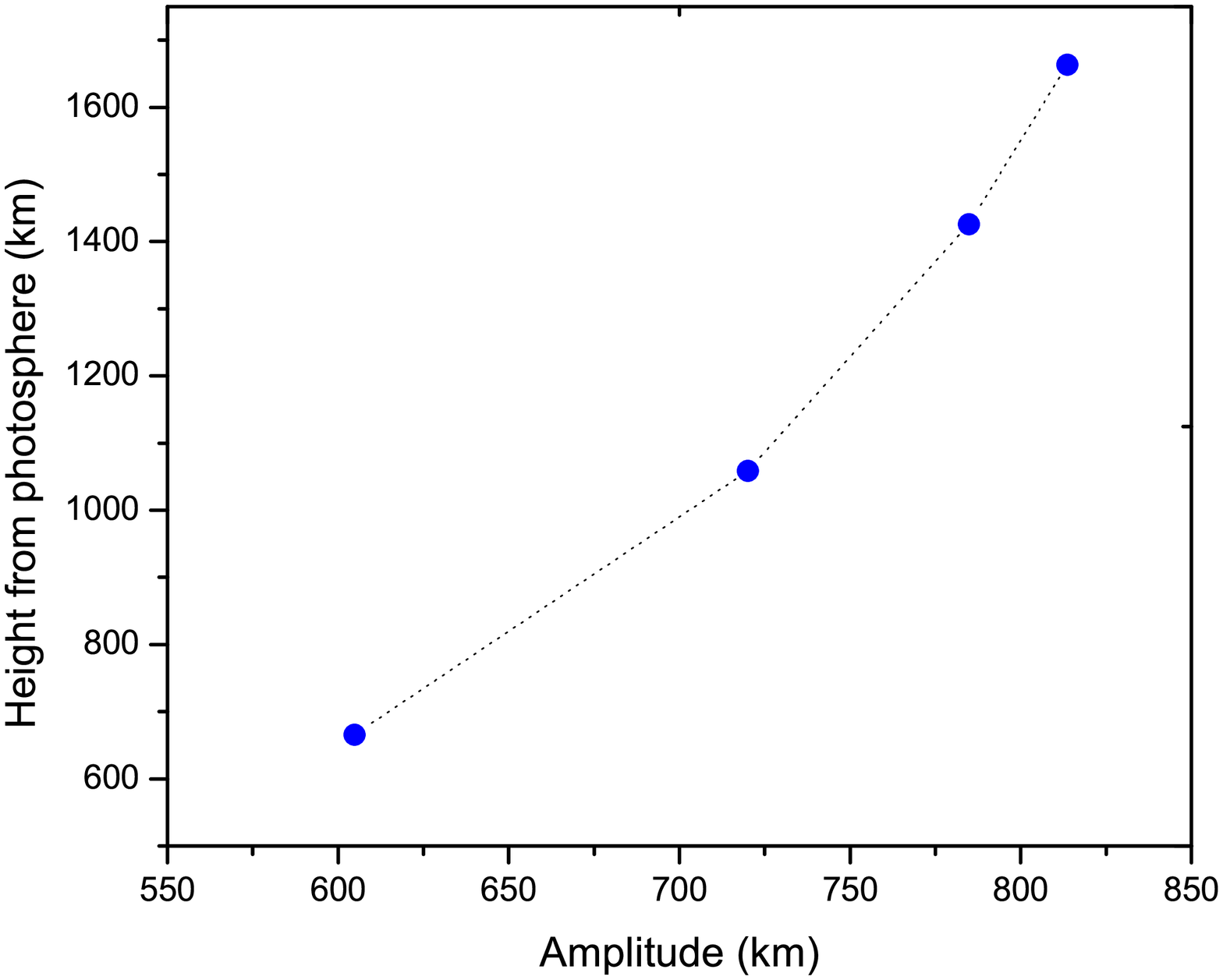}
\includegraphics[width=8cm]{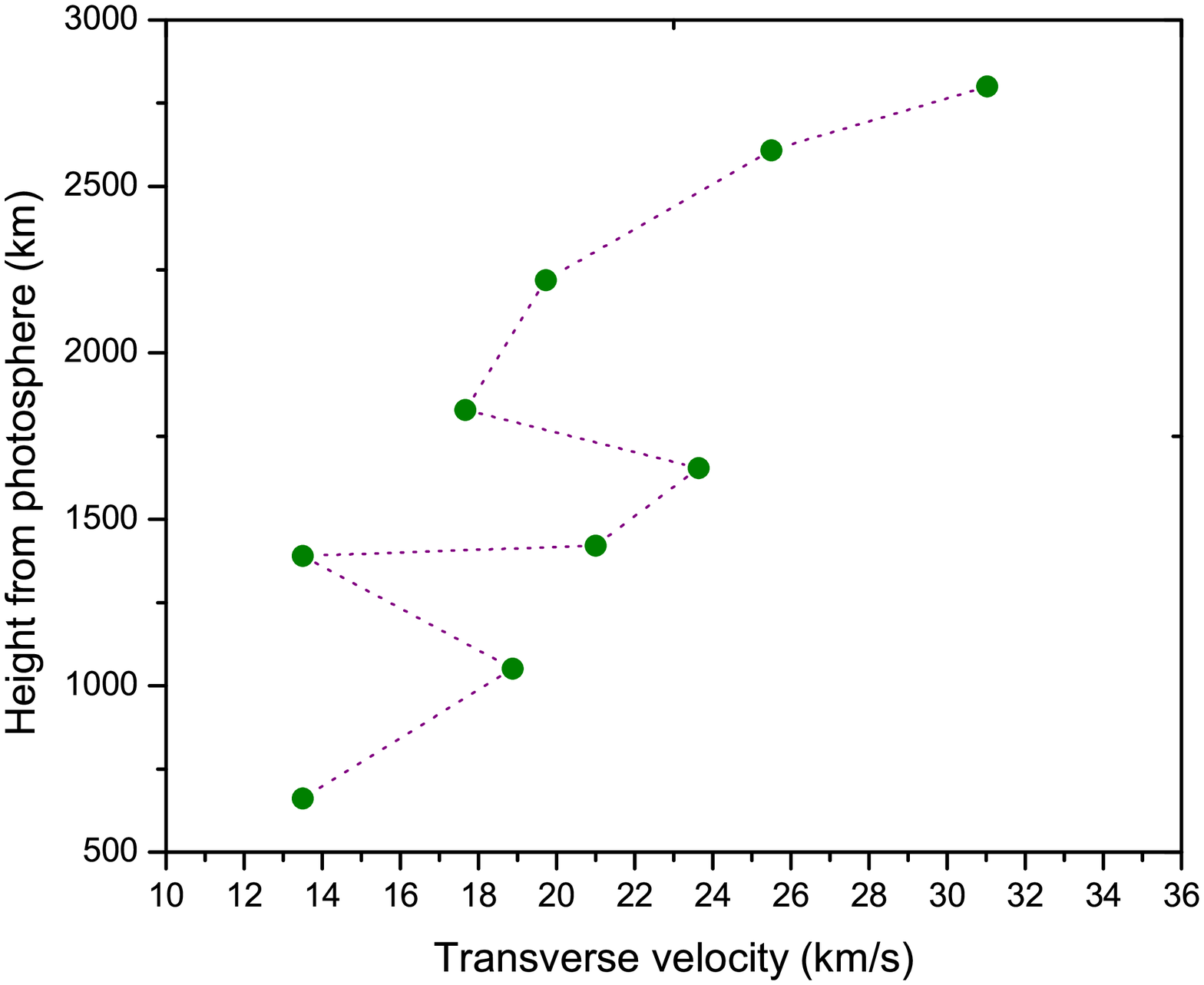}
\caption{The height variations of period, amplitude, and transverse velocity of the spicule axis oscillations are
presented from top to bottom, respectively.\label{fig4}}
\end{figure}

We determined the periods in each height and plotted them with respect to height in the top panel of
Figure~\ref{fig4}. The mean oscillation period in these heights is estimated as $\sim\!\!\! 175$ s.
The oscillation periods are increasing linearly with height.
Kink waves often exhibit only one dominant frequency in spicules \citep{Verth2011}.
\citet{Verth2011} have used the high spatio-temporal Hinode/SOT observations to estimate the vertical gradient in the  magnetic field
and plasma density within a spicule using the change in velocity and phase speed with the height.
Using the observations recorded from Coronal Multi-Channel Polarimeter (CoMP), it has been reported that the broader range of the
frequency of kink waves show a frequency-dependent damping length. This may be most likely due to the perpendicular gradient in the
equilibrium Alfv\'{e}n speed with respect to the magnetic field through the process
of resonant absorption \citep{Verth2010,Goossens2011}.
This mechanism is at work in the range of the chromospheric or coronal wave-guides of different
spatial scales where propagating or standing kink waves are excited, and recently been reported
in the coronal observations \citep{Srivastava2013}. It has been reported that this particular process
causes solar magnetic tubes to act as a natural low-pass filters for the kink waves \citep{Terradas2010}.
The similar kind of frequency filtering is also reported in chromospheric flux tubes for the torsional
Alfv\'{e}n waves \citep{Fedun2011}. Therefore, it is quite possible that the observed period shift
towards the higher period (low frequency) with height within the observed spicule,
is the signature that the spicule is working as a low pass filter and allows only the low frequencies
(higher periods) to propagate towards higher heights.
This provides a magnetoseismology clues that the resonant absorption may be at work in the observed spicule.

The oscillation amplitude is nearly $1$ arc\,sec, but slightly changes with height.
Oscillation amplitude grows with height due to significant decrease
in density, which acts as inertia against oscillations. It should be emphasized that the density scale
height in spicules is $\sim700$ km \citep{Verth2011}.
In bottom panel of Figure~\ref{fig4} we estimated the transversal displacement velocity of spicule axis
with respect to time. The mean transverse velocity is estimated from this figure $\sim\!\!\!18$ km/s.

\begin{figure}
\centering
\includegraphics[width=8cm]{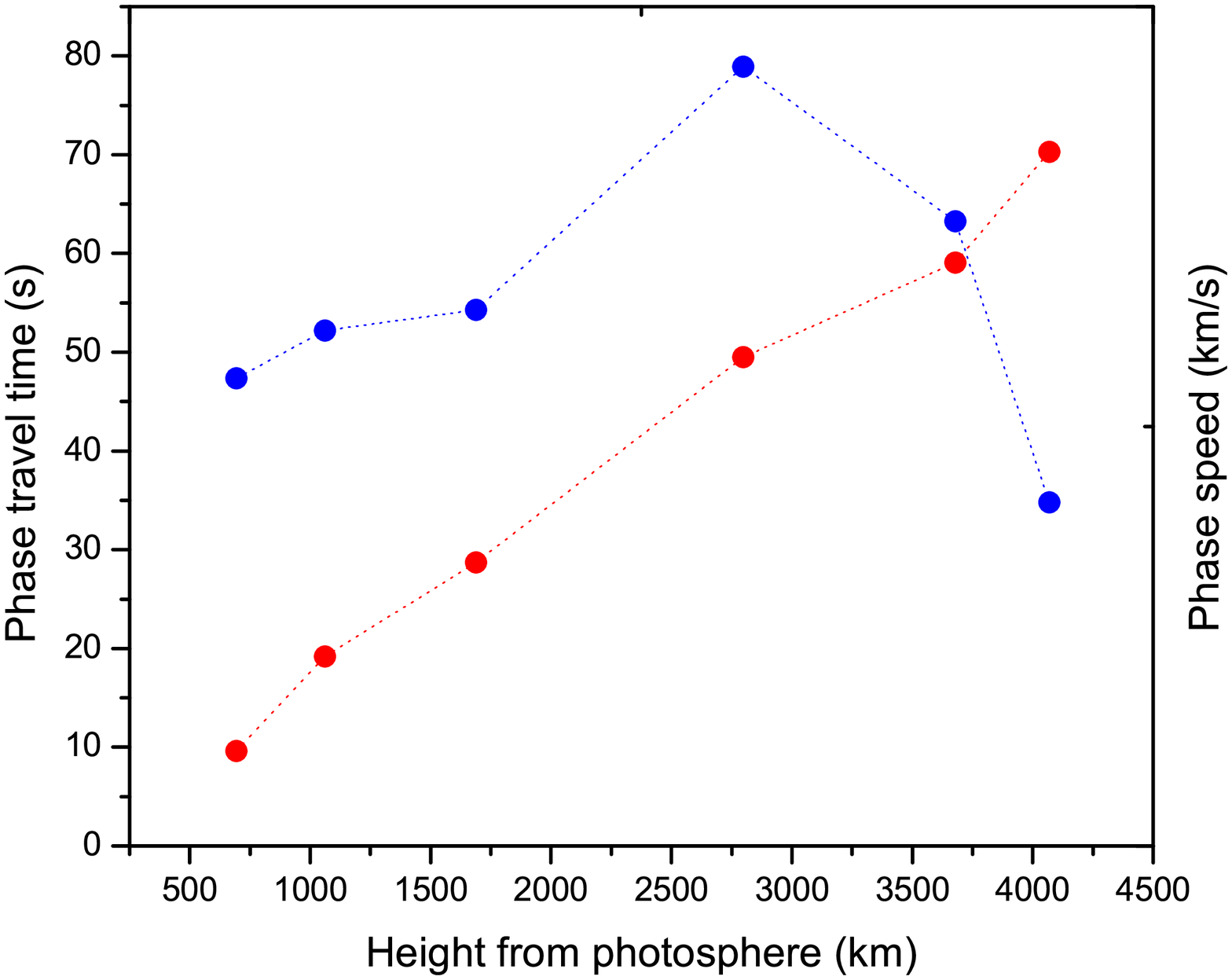}
\includegraphics[width=8cm]{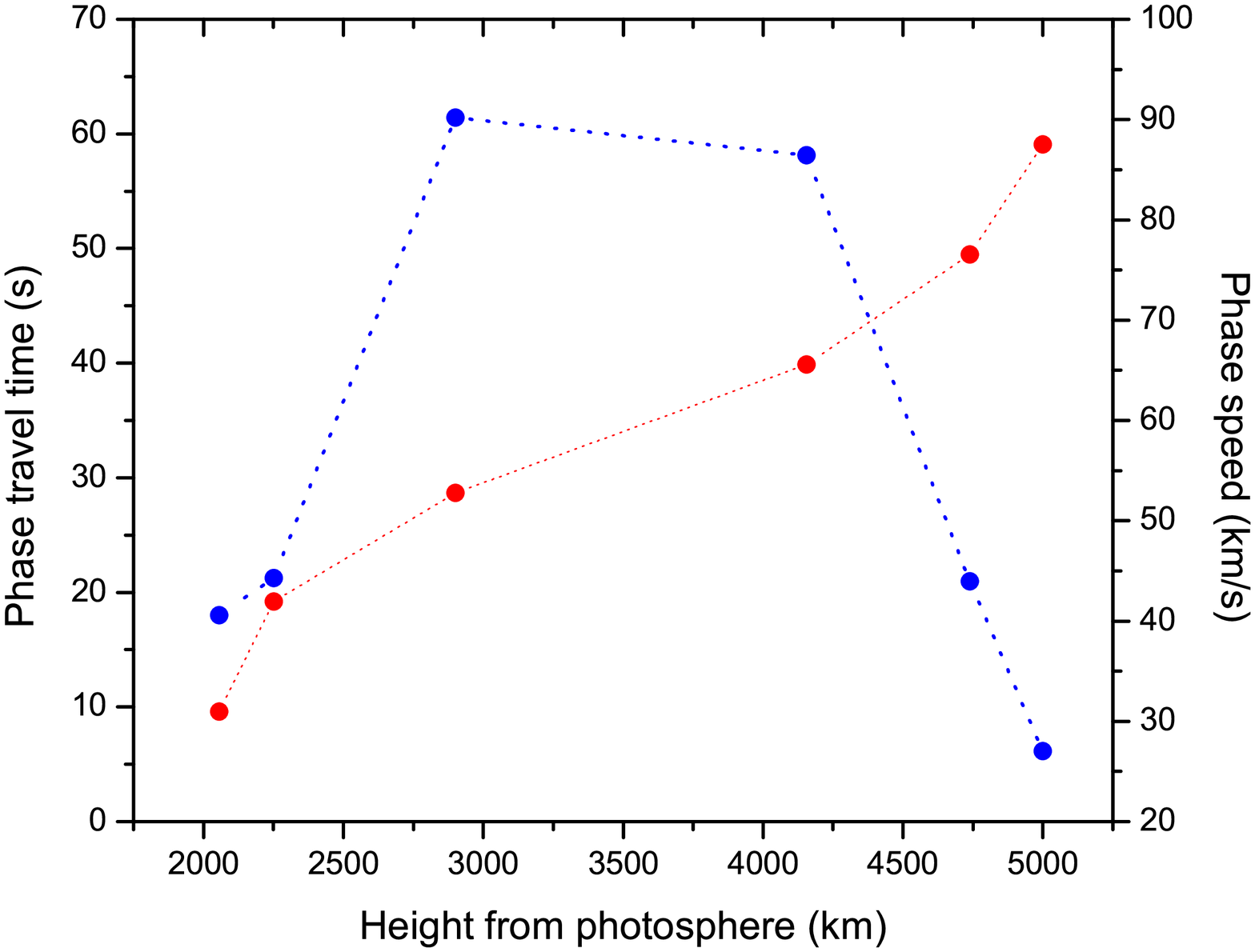}
\includegraphics[width=8cm]{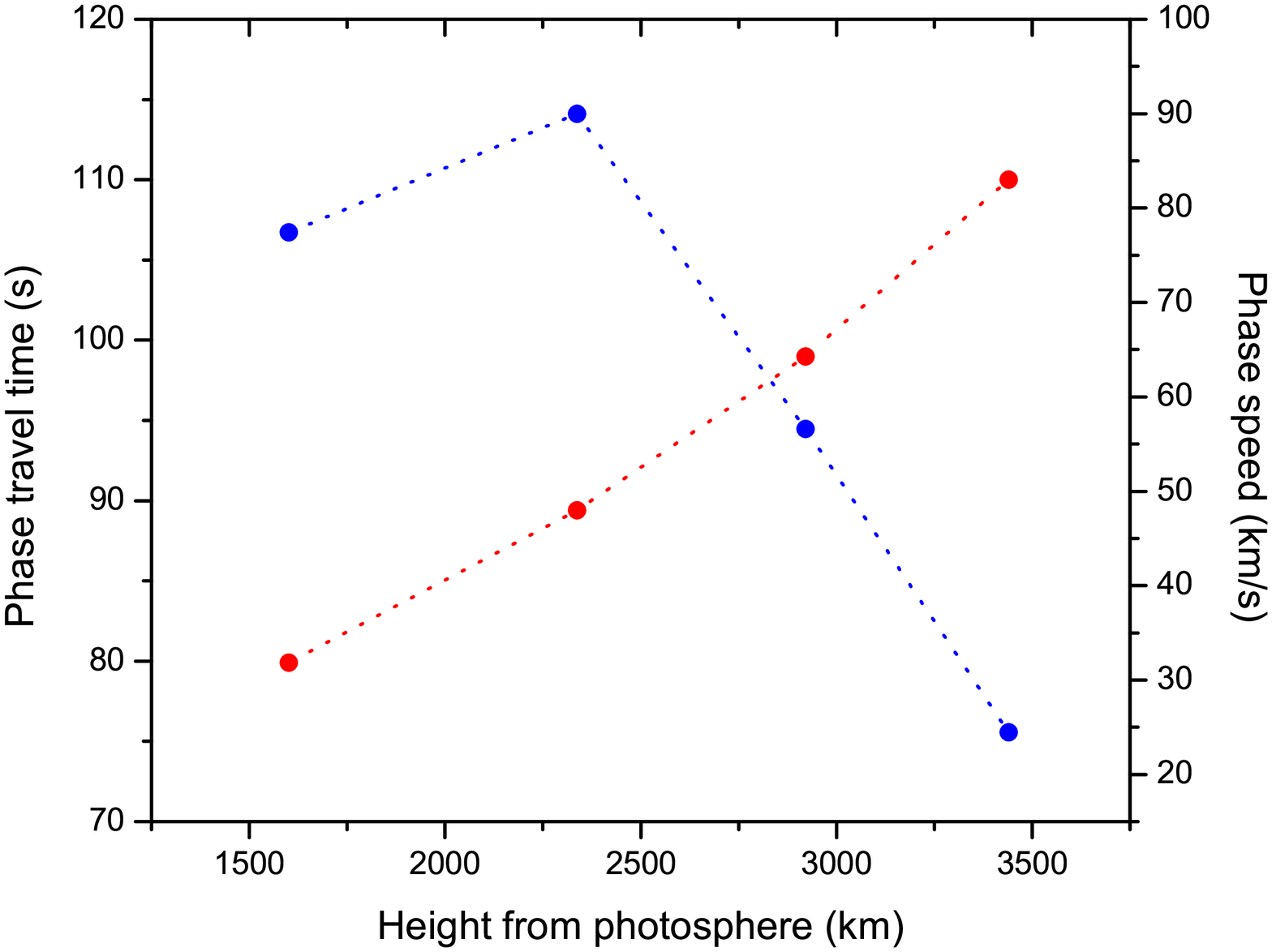}
\caption{The phase travel times and phase speed profiles for three cases of the studied spicule are
showed from top to bottom, respectively. Red plots corresponded to phase travel times and
the blue ones are related to phase speeds profiles. \label{fig5}}
\end{figure}

We presented the phase travel times and phase speed profiles for three cases of the studied spicule in Figure~\ref{fig5}
from top to bottom, respectively. The phase speed is calculated from the first-order derivative of
phase travel time. It begins from $\sim 40$ km/s at lower heights and reaches to the maximum value of $\sim 90$
km/s at $\sim 2500-3000$ km. Then it decreases to the minimum value of $\sim 20-25$ km/s at $\sim 3500-4500$ km.
An increase in phase speed at lower heights may be related to amplification of magnetic field away from reconnection
site. On the other hand, the decreasing behavior at higher heights related to the decrease in mass density and magnetic field strength
in these heights. The mass density and magnetic field variations are studied by \citet{Verth2011} observationally.

The observed quasi-periodic displacement of spicule
axis may be caused due to propagation of kink waves along spicules \citep{he2009,Ebadi2012}.
Transverse oscillation of kink wave may be generated by impact of plasmoid as released from
reconnecting current sheet, or be generated due to reconnection and release of shearing magnetic field.

\section{Conclusion}
\label{sec:concl}
We performed the analysis of Ca \begin{footnotesize}II\end{footnotesize} H-line time series at the solar limb obtained from \emph{Hinode}/SOT in order to uncover the oscillations in the solar spicules.  We concentrate on particular spicule and found that its axis undergos quasi-periodic transverse displacement about a hypothetic line.  The period of the transverse displacement is $\sim\!\!\!175$ s and the mean amplitude is $\sim\!\!1$ arc\,sec.  The same periodicity was found in Doppler shift oscillation by \citet{Ebadi2013,Tem2007,De2007}, so the periodicity is probably common for spicules.

The inverted Y-shaped structures are seen at the spicule base which may correspond
to the magnetic reconnection origin of it. We observed period shift towards the higher period with height within the observed spicule,
which is the signature that the spicule is working as a low pass filter and allows only the low frequencies
to propagate towards higher heights.
We determined the phase travel times and phase speed profiles for three cases of the studied spicule.
Phase speed begins from $\sim 40$ km/s at lower heights and reaches to the maximum value of $\sim 90$
km/s at $\sim 2500-3000$ km. Then it decreases to the minimum value of $\sim 20-25$ km/s at $\sim 3500-4500$ km.
An increase in phase speed at lower heights may be related to amplification of magnetic field away from reconnection
site. On the other hand, the decreasing behavior at higher heights related to the decrease in mass density and magnetic field strength
in these heights.

Therefore, the observed quasi-periodic displacement of spicule axis can be caused due to propagation of kink waves.
The energy flux storied in the oscillation is estimated as $150$ J\,m$^{-2}$\,s$^{-1}$, which is of the order of coronal energy losses in quiet Sun regions.

\acknowledgments
The authors are grateful to the \mbox{\emph{Hinode}} Team for providing the observational data.
\mbox{\emph{Hinode\/}} is a Japanese mission developed and lunched by ISAS/JAXA,
with NAOJ as domestic partner and NASA and STFC(UK) as international partners.
Image processing Mad-Max program was provided by Prof.~O.~Koutchmy.

\makeatletter
\let\clear@thebibliography@page=\relax
\makeatother

\end{document}